\documentclass[prb,twocolumn,english,showpacs,superscriptaddress]{revtex4}
\usepackage[latin9]{inputenc}
\usepackage{amsmath,amssymb,amsfonts}
\usepackage{CJK}
\usepackage{bm}
\usepackage{tikz}
\usepackage{slashed}

\begin{document}

\title{
General response theory of topologically stable Fermi points and its implications for disordered cases
}

\author{Y. X. Zhao}
\email[]{yuxinphy@hku.hk}

\author{Z. D. Wang}
\email[]{zwang@hku.hk}

\affiliation{Department of Physics and Center of Theoretical and Computational Physics, The University of Hong Kong, Pokfulam Road, Hong Kong, China}

\date{\today}

\begin{abstract}
We develop a general response theory of gapless Fermi points with nontrivial topological charges for gauge and nonlinear sigma fields, which asserts that the  topological character of the Fermi points is embodied as the terms with discrete coefficients proportional to the corresponding topological charges. Applying the theory to the effective non-linear sigma models for topological Fermi points with disorders
in the framework of replica approach, we derive rigorously the Wess-Zumino terms with the topological charges being their levels in the two complex symmetry classes of A and AIII. 
Intriguingly, two nontrivial examples of quadratic Fermi points with the topological charge `2' are respectively illustrated for the classes A and AIII. 
We also address a qualitative connection of topological charges of Fermi points in the real symmetry classes to the topological terms in the non-linear sigma models, based on the one-to-one classification correspondence.\end{abstract}

\maketitle

\section{Introduction}
Recently, topological semimetals have been among the hottest topics in condensed matter physics, including graphene, Weyl semimetals, and the surface states of topological insulators and superconductors \cite{Graphene, WSM1,WSM2,Hasan-1,Hasan-2,Kane-RMP,XLQi-RMP}. This may be attributed to not only their potential applications based on their exotic transport properties, but also  broad interests of anomalous currents in condensed matter physics, quantum field theory anomalies, and topological characters of these gapless modes \cite{WSM-response,Response1,Response2,Zhao-Wang-WSM}. As is known, gapless Fermi points in topological semimetals can be classified by their topological charges with respect to their symmetries \cite{FS-classification,TI-FS,FP-1,FP-2,FP-3}. Although various investigations have been made on phenomena and classifications of these topological points, implications of the topological charges to quantum field theories are still badly awaited to be explored, which is obviously of fundamental importance and interdisciplinary interest.
Mainly motivated by this, here we first establish a general quantitative connection between the topological character of  topological Fermi points and the topological terms with non-perturbative discrete  coefficients in its effective response theory, being coupled to external sources that may be gauge or sigma fields. These topological terms  correspond usually to anomalous transport properties of these Fermi points, and are related to quantum field theory anomalies \cite{Peskin}.
For instance, in Weyl semimetals and the A-phase of 3He, 
 there are Weyl points with nontrivial Chern numbers as their topological charges,
 which leads to the abelian chiral anomaly in the $U(1)$-response theory with anomalous currents \cite{Volovik-book,Chiral-anomaly-1,Chiral-anomaly-2,Fujikawa,Peskin,Qi-Liu,WSM-response}. We then make important applications of our general theory to the response theories 
 for disordered Fermi points with nontrivial topological charges using the replica approach \cite{Replica-I,Replica-II, Pruisken}.
 On one hand, various disorders are ubiquitous in most real materials, while, on the other hand, as we will see, anomalies of non-abelian gauge theories are naturally related to such condensed matter problems \cite{Fred-I,Supersymmetry-Z2,Ryu-Z2,Polyakov,WZ,YSWu,Witten-WZ,Nakahara}. As our main results, the integer topological charges ($\nu$) of Fermi points in the complex symmetry classes of A and AIII in the Altland-Zirnbauer (AZ) classification \cite{AZClasses,AZClasses-2} is rigorously shown to lead to the Wess-Zumino terms (WZ terms) at level $\nu$ \cite{WZ,Witten-WZ}. For the class A, the emergence of the WZ terms is related to the parity anomaly in odd dimensions \cite{Parity-anomaly}; while  for the class AIII, it is associated with the non-abelian anomaly in even dimensions \cite{WZ,YSWu,Nakahara}. Finally, we also address the qualitative relationship of topological charges in the eight real AZ classes to the topological terms with discrete coefficients. While such topological terms have been indicated to have important consequences in the nonlinear sigma models(NL$\sigma$Ms) \cite{AIII-2d-1,AIII-2d-2,Fred-II,Fred-III} and argued to appear in various systems \cite{Fred-I,Fred-II,Fred-III,Schnyder-classification}, such as the disordered boundaries of topological insulators~\cite{Schnyder-classification},
 we here identify unambiguously their origins for the Fermi points as the corresponding topological charges.

\section{Stable equivalence and universal responses} 
Let $\mathcal{O}(k)$ be a set of (hermitian) bounded operators in a Hilbert space, which are parametrized by $k$ in a Euclidean space and can be regarded as a mapping from $k$ space to linear transformations. We assume that there are gapless points inside a finite region with zero eigenvalue, namely the spectrum gap opens far away from the origin of the $k$ space. For instance, with regard to  the band structure of a solid, the operators are just Hamiltonians $\mathcal{H}(k)$ in the Brillouin zone, while for a quantum field theory with only quadratic terms in the Euclidean formulation, the operators are the Lagrangian density $\mathcal{L}(k)$. Let $\mathcal{O}_1(k)$ and $\mathcal{O}_2(k)$ are two such operator distributions, whose dimensions may be different. The \textit{stable} equivalence between $\mathcal{O}_1$ and $\mathcal{O}_2$  is defined in the following way. After adding arbitrary numbers of trivial gapped bands to $\mathcal{O}_j$, namely $$\tilde{\mathcal{O}}_j(k)=\mathcal{O}_j(k)\oplus p_+\cdots\oplus p_+\oplus p_{-}\cdots\oplus p_{-}$$ with $p_{\pm}=\pm1$, if $\tilde{\mathcal{O}}_1(k)$ and $\tilde{\mathcal{O}}_2(k)$ can be smoothly deformed to each other without closing the gap far away from the origin of the $k$ space, $\mathcal{O}_1(k)$ and $\mathcal{O}_2(k)$ are \textit{stably} equivalent, $\mathcal{O}_1(k)\approx \mathcal{O}_2(k)$ \cite{Karoubi,Kitaev-classification,Jeffrey}. Note that we are actually defining asymptotic stable topological configurations for a operator distribution.   If a set of symmetries of the operator distributions are required for the smooth deformations, then the two $\mathcal{O}$'s are said to be \textit{stably} equivalent under the symmetries. For such an $\mathcal{O}(k)$, we can choose an $S^{d-1}$ in the $d$-dimensional $k$ space far away from the gapless region, on which $\mathcal{O}(k)$ is gapped, and  define a topological charge valued in an abelian group (for instance, $\mathbb{Z}$ or $\mathbb{Z}_2$)  by $K$-theory \cite{Karoubi,Kitaev-classification,FS-classification}. Two $\mathcal{O}$'s of the same topological charge are stably equivalent.  For the ten AZ symmetry classes, distinct collections of stably equivalent gapless Hamiltonians in every AZ class are identified by their topological charges that are symmetry-related topological invariants formulated on the gapped spheres enclosin of the gapless regions from the transverse dimensions in $k$ space \cite{FS-classification,TI-FS}.

Now we are ready to develop in general a topological response theory of a quadratic quantum field theory in the Euclidean formulation, $S=\int dk \bar{\psi} (k)\mathcal{L}(k)\psi(k)$ where $\mathcal{L}(k)$ has nontrivial topological charge $\nu$, which is coupled to an external source field $Q(x)$ corresponding to some long-distance degree of freedom compared with the ultraviolet cutoff of the theory. For a gauge theory, $Q(x)$ is just the gauge field, while for a sigma model, $Q(x)$ is in some target manifold, for instance $S^n$, $SU(N)$ or some more complicated ones we will meet soon.
The low-energy effective theory is given by integrating over the fermionic fields,
\begin{equation}
-\ln\frac{\mathrm{Det}[-\mathcal{L}(k, Q(x))]}{\mathrm{Det\mathcal{L}(k)}}=\sum_{j}\lambda_j S^{(j)}[Q],\label{Eff}
\end{equation}
namely $S_{eff}[Q]=\sum_{j}\lambda_j S^{(j)}[Q]$, where $\lambda_j$ is the coefficient of the term $S^{(j)}$, which is a product of $Q$s and derivatives of $Q$. We consider the family of terms $S_{top}[Q]$ that are originated from the stable topological property of $\mathcal{L}(k)$, and therefore are invariant under the smooth deformations of $\mathcal{L}(k)$ specified by the stable equivalence in the previous paragraph, noting that adding trivial bands such as $p_{\pm}$ does not affect the effective theory of Eq.(\ref{Eff}).  In general $\nu\in\oplus_\alpha\mathbb{Z}_{n_\alpha}$, where $\mathbb{Z}_{n_\alpha}$ is just the abelian group of integers modulo $n_\alpha$ with the convention that $\mathbb{Z}_{n_\alpha=\infty}=\mathbb{Z}$. Since  $\mathcal{L}$ can always be diagonally blocked as $\mathcal{L}=\oplus_\alpha\mathcal{L}_\alpha$ according to the group $\nu\in\oplus_\alpha\mathbb{Z}_{n_\alpha}$ with $\mathcal{L}$ corresponding to $\mathbb{Z}_{n_\alpha}$ after smooth deformations, we find that $$S_{top}[Q]=\sum_\alpha \lambda_\alpha S_{top}^{\alpha}[Q],$$ noting that the direct sum is translated to addition from the left side to the right of Eq.(\ref{Eff}). Accordingly, it is expected that $\lambda_{\alpha}$ is a function of $\nu_\alpha\in\mathbb{Z}_{n_\alpha}$, i.e., $\lambda_\alpha(\nu_\alpha)$, which is actually  proportional to $\nu_\alpha$: $\lambda_\alpha(\nu_\alpha)=\kappa\nu_\alpha$ with $\kappa$ being a constant. This is because that  an $\mathcal{O}(k)$ of topological charge $\nu\in\mathbb{Z}_n$ can always be smoothly deformed to be a multiple of $\nu$ identical gapless $\mathcal{O}_1$'s with each having a unit topological charge, $\mathcal{O}(k)\approx \oplus_{j=1}^{\nu} \mathcal{O}_1(k)$, after adding sufficient number of trivial bands. Thus, as a general theoretical observation, it is found that
\begin{equation}
S_{top}[Q]=\sum_\alpha\nu_\alpha S_{top}^\alpha[Q], \label{Top-term}
\end{equation}
where each $\kappa$ has been absorbed into $S_{top}^\alpha$ for convenience. Eq.(\ref{Top-term}) is a general quantitative relation between the topological charges of Fermi points and topological terms of non-perturbative discrete coefficients, serving as one of our main results.

We now make several comments on Eq.(\ref{Top-term}). First, each generator for the $K$-group may lead to a topological term of Eq.(\ref{Top-term}) with a discretely valued coupling constant being proportional to the corresponding topological charge. Such terms usually have some topological meanings. For instance, in a gauge theory candidates may be Chern-simons terms for non-abelian groups with the quantization of coupling constant given by the gauge invariance, or Chern characters characterizing different topological classes of gauge equivalent configurations classified by an abelian group with twisted components \cite{Volovik-book,Nakahara,Fujikawa}. For an NL$\sigma$M, WZ and $\mathbb{Z}_n$-$\theta$ terms are naturally such terms, with the quantization of coupling constants given by definitions \cite{Schnyder-classification,Supersymmetry-Z2,Ryu-Z2}. Secondly, the regularization of the quantum field theory in Eq.(\ref{Eff}) is required to respect the topological property of $\mathcal{L}$ specified by the stable equivalence. A normal Pauli-Villa regulator is applicable, since it adds only very massive propagators corresponding to trivial pairs of gapped bands that are allowed by stable topological equivalence.

\section{Wess-Zumino terms of the class A} 
Now we apply Eq.(\ref{Top-term}) to the NL$\sigma$Ms of  topological Fermi points with disorders through the replica trick, starting with the class A that possesses no any discrete symmetry. As is known, nontrivial topological points in the class A can exist only in odd dimensions, $d=2n+1$, due to the Bott periodicity \cite{FS-classification,Karoubi}. The corresponding topological charge $\nu_{A}$ of a Fermi point is given as the Chern number of the Berry bundle of occupied bands on the gapped $2n$-dimensional sphere enclosing the Fermi point in $k$ space\cite{Volovik-book}. For example, a formula for calculating the topological charge of the Weyl point $\mathcal{H}_{W}=\sigma\cdot \mathbf{k}$ is given by
\begin{equation}
\nu_{A}[G_{W}]=\frac{1}{24\pi^2}\int_{S^3}\mathrm{tr}(G_{W}dG_{W}^{-1}(\omega,k))^3 ,
\end{equation}
where $G_{W}=1/(i\omega-\mathcal{H}_{W})$ is the imaginary Green's function, and the $S^3$ is a three-dimensional sphere chosen in $(\omega,k)$ space enclosing the gapless points.
In the class A, the sigma field $$Q\in BU=U(2N)/(U(N)\times U(N))$$ can describe the low-energy degrees of freedom near a saddle point of  the mean-field theory in the standard replica method, where $N$ denotes the number of replicated systems. The topological term in Eq.(\ref{Top-term}) for a gapless $\mathcal{H}_{A}(k)$ with an integer topological charge $\nu_{A}\in\mathbb{Z}$ is the WZ term at level $\nu_{A}$, which  can be written as
\begin{equation}
S_{WZ}^{A}[Q]=\nu_{A}C_{d}\int_{D^{d+1}} \mathrm{tr}\tilde{Q}(d\tilde{Q})^{d+1},\label{WZ-A}
\end{equation}
based on a rigorous analysis to be detailed from the next paragraph, where $$C_d=\frac{-n!}{(2n)!(2\pi i)^n 2^{2n+3}},$$
and $\tilde{Q}(x,\tau)$ is a continuous extension of $Q(x)$ along the parameter $\tau\in[0,1]$ with $\tilde{Q}(x,1)=Q(x)$ and $\tilde{Q}(x,0)$ being constant. Since the homotopy group $\pi_{2n+1}(BU)=0$, the extension is always possible. Accordingly, the original real space is extended to the $(d+1)$-dimensional disk $D^{d+1}$, whose boundary $S^{d}$ is assumed to be the original real space after compactification.  The difference of the values of Eq.(\ref{WZ-A}) for two extensions is $2\pi i \nu_{A} N_A$ with the integer $N_A$ being the winding number difference of the two extensions recalling that $\pi_{2n+2}(BU)\cong\mathbb{Z}$, which justifies that the WZ-term is well defined \cite{Witten-WZ}. In particular, the coefficient of a WZ term can only take discrete values labelled by its level $m_A$, which is perfectly in consistence with a fact that the topological charge as a topological invariant is an integer, considering that $m_A=\nu_{A}$ in Eq.(\ref{WZ-A}). It is noted that Eq.(\ref{WZ-A}) can be argued from the boundary-bulk correspondence of a disordered $(2n+2)$-dimensional Chern insulator \cite{Zhao-Wang-WSM}.

To prove Eq.(\ref{WZ-A}), as what we discussed above Eq.(\ref{Top-term}), it is sufficient to consider merely the case for the unit topological charge $\nu_{A}=1$, which can be realized by Dirac type Hamiltonian, $$\mathcal{H}(k)=\sum_{j=1}^{2n+1} k_j\Gamma_{j}^{(2n+1)}$$ with $\Gamma_j^{(2n+1)}$ being $2^{n}\times 2^{n}$ Dirac matrices, satisfying \cite{TI-FS}$$\{\Gamma_i,\Gamma_j\}=2\delta_{ij}.$$ We adopt the Hamiltonian of a Weyl point $\mathcal{H}(k)=\mathbf{k}\cdot\sigma$ with $d=3$ to exemplify the proof, and it is straightforward to see its validity in any odd dimensions. The sigma field can be explicitly expressed as $Q=T\tau_3 T^{-1}$, where $T(x)\in U(2N)$ and $\tau_3$ is the third Pauli matrix acting on the retarded-advanced space. Before evaluating the functional determinant in Eq.(\ref{Eff}), it is useful to express the WZ action in terms of $T$. It is straightforward, although tedious, to check that
\begin{equation*}
\mathrm{tr}\tilde{Q}(d\tilde{Q})^4=\frac{8}{3}d\mathrm{tr}(T^{-1}dT\tau_3)^3-8d\mathrm{tr}\tau_3(T^{-1}dT)^3,
\end{equation*}
which implies that the WZ action for unit $\nu_{A}$, in terms of $T$, can be written explicitly in the original real space $S^3$ as
\begin{equation*}
S_{WZ}^{A}=\frac{i\nu_{A}}{48\pi}\int_{S^3} \mathrm{tr}(T^{-1}dT\tau_3)^3-3\mathrm{tr}\tau_3(T^{-1}dT)^3.
\end{equation*}
For brevity, we define the projectors $P_{\pm}=(1\pm\tau_3)/2$ for the advanced and retarded spaces and $A=T^{-1}dT$, accordingly the action is translated to be
\begin{equation}
S_{WZ}^{A}=\frac{i\nu_{A}}{2}(S_{CS}[AP_{+}]-S_{CS}[AP_{-}]), \label{CS-form}
\end{equation}
where $$S_{CS}[A]=\frac{1}{4\pi}\int_{S^3}\mathrm{tr}[AdA+\frac{2}{3}(A)^3]$$ is the Chern-Simons (CS) term. The Chern-Simons expression has been derived in studying a single disordered Weyl point for a Weyl semimetal,  seemingly without realizing it to be actually a WZ term \cite{Altland-WSM}.

\begin{figure}
\includegraphics[scale=0.25]{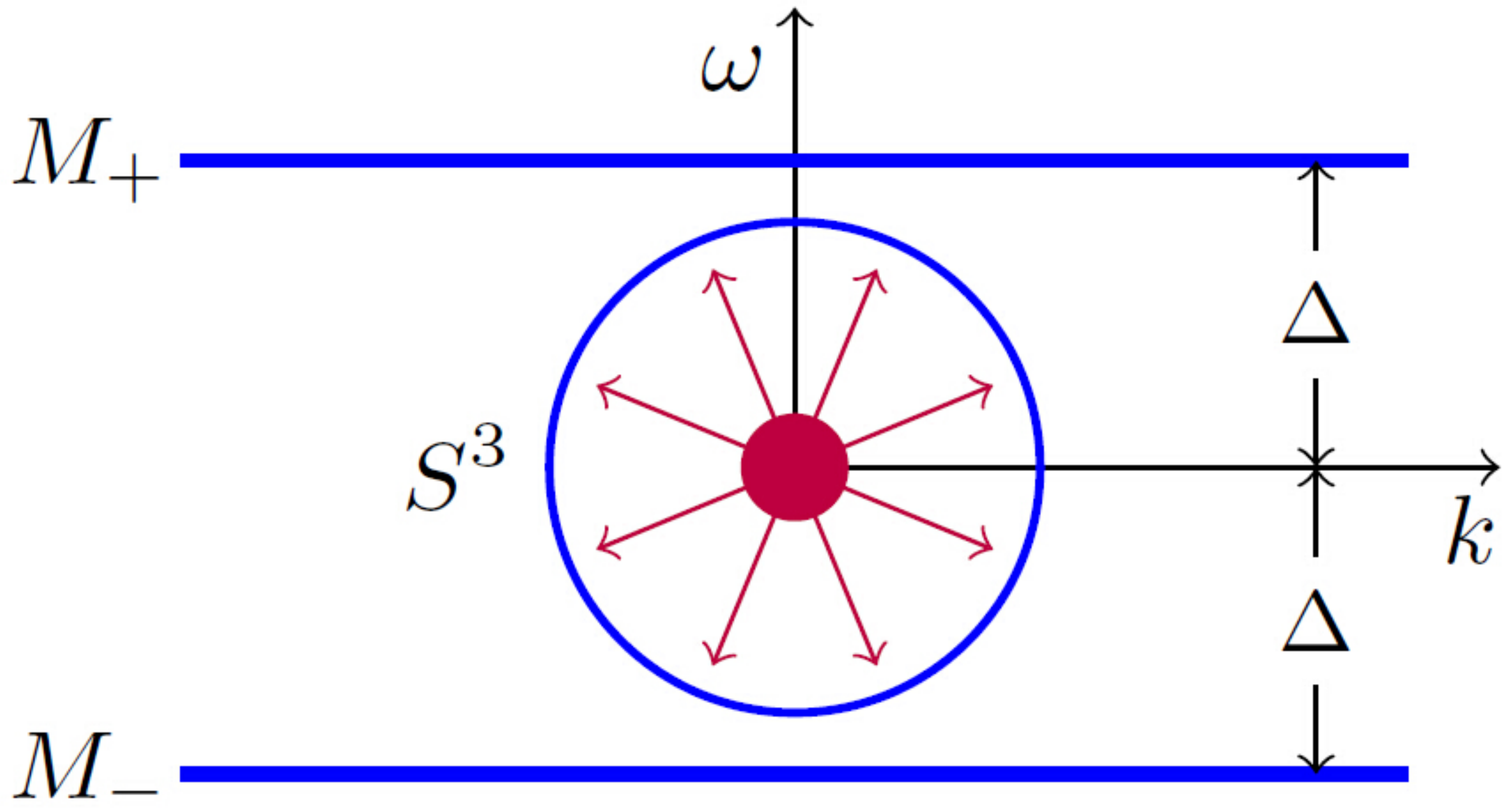}
\caption{The $(\omega,k)$ space with a Weyl point at the centre. $M_{\pm}$ is the $k$ space of $G_{\pm}$ and $S^3$ is chosen to enclose the Weyl point. \label{Weyl-point}}
\end{figure}

Now our aim is to deduce Eq.(\ref{CS-form}) from the functional determinant $\mathrm{Det}(-\slashed k+i\Delta Q)$ with $\slashed k=\sigma\cdot\mathbf{k}$, recalling Eq.(\ref{Eff}) with $g=i\Delta$ in this case. After a unitary transformation we have
\begin{equation*}
\begin{split}
\mathrm{Det}T^{-1}(-\slashed k+i\Delta Q)T&\sim\mathrm{Det}(-\slashed k+i\slashed A+i\Delta \tau_3)\\
&=\mathrm{Det}\left(1+G(\Delta,k)i\slashed A\right )
\end{split}
\end{equation*}
with $$G(\Delta,k)=\frac{1}{i\Delta \tau_3-\slashed k}$$ as a propagator and $i\slashed A$ as a vertex. The regulator may lead to additional modifications \cite{Parity-anomaly}, which do not affect the present discussions. So the effective theory is given by a summation of one-loop Feynman diagrams, $$S_{eff}[Q]=\sum_{n=1}^{\infty}\frac{(-1)^n}{n}\mathrm{Tr}(G(\Delta,k)i\slashed A)^n.$$ It is well-known that the CS terms, Eq.(\ref{CS-form}), that we are searching for are related to the parity anomaly in odd dimensions \cite{Parity-anomaly}, which provides us a clue to derive them from the Feynman diagrams, analogous to the derivation of the CS term for electromagnetic response of a $(2+1)$-dimensional Chern insulator\cite{Volovik-book,Kane-RMP,XLQi-RMP}, but with extra complications and new interpretations.   Considering that $A(q)$ encodes low-energy freedoms of small momentum $q$, a term with the same form of Eq.(\ref{CS-form}) can be collected from the two-vertex loop and three-vertex one,
\begin{equation}
S=iN[G_{+}]S_{CS}[AP_{+}]+iN[G_{-}]S_{CS}[AP_{-}],\label{CS-temp}
\end{equation}
where $G_{\pm}=1/(\pm i\Delta-\slashed k)$, and $$N[g]=\frac{1}{24\pi^2}\int_{M}\mathrm{tr}(gdg^{-1})^3$$ is an integration over the whole momentum space $M$. To see the topological origin of Eq.(\ref{CS-temp}), let us regard $\pm\Delta$ as the given values of $\omega$, and therefore viewed in $(\omega,k)$ space the momentum spaces of $G_{\pm}$ are two parallel three-dimensional surfaces $M_{\pm}$ sandwiching the Weyl point as shown in Fig.(\ref{Weyl-point}). Since $M_{\pm}$ are two infinite parallel surfaces, the union of them are topologically equivalent to a three-dimensional sphere $S^3$ enclosing the gapless point as shown in Fig.(\ref{Weyl-point}), namely $$\nu_{A}[G_{W}]=N[G_{+}]-N[G_{-}]$$ with orientations of surfaces being considered. Since $$|N[G_{+}]|=|N[G_{-}]|$$ due to $\omega=\pm\Delta$, we find
\begin{equation}
N[G_{+}]=-N[G_{-}]=\nu_A[G_W]/2,
\end{equation}
and it follows that Eq.(\ref{CS-temp}) is just Eq.(\ref{CS-form}), recalling that $\nu_A[G_W]=1$. In conclusion, we have identified the topological terms of Eq.(\ref{Top-term}) for the class A is just the WZ terms, Eq.(\ref{WZ-A}), because the generalization to other odd dimensions is obvious.

As an intriguing example, we consider a quadratic Fermi point, $$\mathcal{H}_{A}(k)=(k_x^2-k_y^2)\sigma_1+2k_xk_y\sigma_2+k_z\sigma_3.$$ As shown in Appendix \ref{A-Inter}, a continuous interpolation can be constructed after adding two trivial bands, so that it is smoothly deformed to be a doublet of $\sigma\cdot\mathbf{k}$. Thus, its NL$\sigma$M for disorders has the term of Eq.(\ref{WZ-A}) with $\nu_A=2$ as a nontrivial application of our result, although direct derivation of the response theory in the quadratic case is extremely difficult.


\section{Wess-Zumino terms of the class AIII}
A Hamiltonian $\mathcal{H}_C(k)$ in the class AIII has a chiral symmetry, namely there is a unitary matrix $\Gamma$ anti-commuting with $\mathcal{H}_C(k)$, $\{\mathcal{H}_C,\Gamma\}=0$. Without loss of generality, we assume that $\Gamma=\mathrm{diag}(1,-1)$, and accordingly
\[
\mathcal{H}_C=\begin{pmatrix}0 & \mathcal{H}_{-}\\
\mathcal{H}_{+}& 0\end{pmatrix},
\]
$\psi=(\psi_{+},\psi_{-})^T$ and $\bar{\psi}=(\bar{\psi}_{-},\bar{\psi}_{+})$. Randomness respecting the chiral symmetry leads to $Q=\mathrm{diag}(M,M^{-1})$ with $M\in U(N)$.  Fermi points of $\mathcal{H}_C(k)$ in the class AIII of nontrivial topological charge $\nu_C$ can exist only in even dimensions $d=2n$ \cite{FS-classification}, and the topological term in Eq.(\ref{Top-term}) for $\mathcal{H}_C$ is the WZ term of $U(N)$ at level $\nu_C$, which is given by
\begin{equation}
S_{WZ}^{AIII}[M]=\nu_c C'_d\int_{D^{d+1}}\mathrm{tr}(\tilde{M}^{-1}d\tilde{M})^{2n+1},\label{WZ-AIII}
\end{equation}
after rigorous derivations to be presented later, where $$C'_d=-\frac{n!}{(2n+1)!(2\pi i)^n},$$ and $\tilde{M}(x,\tau)$ is the continuous extension of $M(x)$ through $\tau\in[0,1]$ with $\tilde{M}(x,1)=M(x)$ and $\tilde{M}(x,0)$ being constant. The extension independence of Eq.(\ref{WZ-AIII}) has the same reason as that of Eq.(\ref{WZ-A}).

As discussed above Eq.(\ref{Top-term}), to derive the WZ term of Eq.(\ref{WZ-AIII}), it is sufficient to work out $$\mathcal{H}^D=\sum_{j=1}^{2n}\Gamma^{(2n+1)}_jk_j$$ ($\Gamma=\Gamma^{(2n+1)}_{2n+1}$) with unit topological charge~\cite{TI-FS}. Let us use the four-dimensional case $\mathcal{H}^{D}(k)=\slashed k$ with $\slashed k=\sum_{j=1}^{4}\Gamma^{(5)}_jk_j$ to exemplify the proof. First the Lagrangian is
\begin{equation*}
\mathcal{L}=-\bar{\psi}_{+}\slashed k_{+}\psi_{+}-\bar{\psi}_{-}\slashed k_{-}\psi_{-}+g\bar{\psi}_{-}M\psi_{+}+g\bar{\psi}_{+}M^{-1}\psi_{-},
\end{equation*}
with $\slashed k_{\pm}=\slashed k(1\pm \Gamma_5)/2$. Since $\pi_4(U(N))=0$, $M(x)=e^{im(x)}$, where $m(x)\in u(N)$ with $u(N)$ being the Lie algebra of $U(N)$. Let us introduce a series of fields parametrized by $\tau\in[0,1]$, $$\psi_{+}(\tau)=\tilde{M}(\tau)\psi_{+},\quad \bar{\psi}_{+}(\tau)=\bar{\psi}_{+}\tilde{M}(-\tau).$$where $\tilde{M}(\tau)=e^{i\tau m}.$ We further introduce
\begin{multline*}
\mathcal{L}(\tau)=-\bar{\psi}_{+}(\tau)[\slashed k_{+}-i\slashed A_{+}(\tau)]\psi_{+}(\tau)-\bar{\psi}_{-}\slashed k_{-}\psi_{-}\\
+g\bar{\psi}_{-}\tilde{M}(1-\tau)\psi_{+}(\tau)+g\bar{\psi}_{+}(\tau)\tilde{M}^{-1}(1-\tau)\psi_{-},
\end{multline*}
where $A(\tau)=\tilde{M}(\tau)d\tilde{M}(-\tau)$, and the partition function $$Z(\tau)=\int \mathcal{D}\psi(\tau)\mathcal{D}\bar{\psi}(\tau) e^{-\int\mathcal{L}(\tau)}.$$ Let  $\tau$ vary from $\tau=0$ with $\mathcal{L}(0)=\mathcal{L}$ to $\tau=1$ where the dependence of $\mathcal{L}(1)$ on $M$ is entirely encoded in $A(1)=MdM^{-1}$. Under field variable transformations, 
\begin{equation*}
\psi_{+}(\tau)\rightarrow \psi_{+}(\tau+d\tau),\quad
\bar{\psi}_{+}(\tau)\rightarrow\bar{\psi}_{+}(\tau+d\tau),
\end{equation*}
although $\mathcal{L}(\tau+d\tau)=\mathcal{L}(\tau)$, there exists a nontrivial Jacobian determinant $$J(\alpha,\tau)=1+\frac{\delta J}{\delta \alpha}|_{\alpha=0}\alpha$$ with $\alpha=imd\tau$, noting that $\delta\psi_{+}(\tau)=\alpha\psi_{+}(\tau)$. So $Z(\tau)$ satisfies the differential equation, $$dZ(\tau)=Z(\tau)f(\tau)d\tau$$ with $f(\tau)=\frac{\delta J}{\delta \alpha}|_{\alpha=0}im ,$ which implies $$Z(0)=Z(1)\mathcal{J}$$ with $$\mathcal{J}=\exp(-\int_{0}^{1}f(\tau)d\tau).$$ The Jacobian determinant $J(\alpha,\tau)$ depends on regularization schemes, and under the commonly adopted one for non-abelian anomalies, we have \cite{WZ,YSWu,Nakahara} $$f(\tau)=-\frac{1}{24\pi^2}\int_{S^4} \mathrm{tr}[im\,d(AdA+A^3/2)].$$ Plugging in $A(\tau)=\tilde{M}(\tau)d\tilde{M}(-\tau)$, we have $$f(\tau)=\frac{1}{48\pi^2}\int_{S^4} \mathrm{tr}[\tilde{M}(-\tau)\frac{\partial \tilde{M}(\tau)}{\partial\tau}(\tilde{M}(-\tau)d\tilde{M}(\tau))^4].$$ Thus in the effective action of $M$, there exists the term $\int_0^{1} ds f(s)$ that is explicitly written as
\begin{equation}
S[M]= \frac{1}{240\pi^2}\int_{D^5} \mathrm{tr}(\tilde{M}^{-1}d\tilde{M})^5,\label{4D-WZ-AIII}
\end{equation}
where $D^5$ is naturally given by $S^4\times[0,1]$ with $\tilde{M}(0)=1$ and $\tilde{M}(1)=M$. Eq.(\ref{4D-WZ-AIII}) is exactly Eq.(\ref{WZ-AIII}) when $n=2$.

\subsection{NL$\sigma$M of a 2d AIII Fermi point}
For two dimensions, the Dirac matrices are just Pauli matrices and $\mathcal{H}_{2d}^{D}=\sigma_1 k_x+\sigma_2 k_y$ with a unit topological charge. The existence of WZ terms for $\mathcal{H}_{2d}^{D}$ has been argued by Fendley in the study of a $p$-wave triplet superconductor model in the same spirit of the above non-abelian anomaly \cite{Fred-I}. In this case, the above regularization scheme suffers from a fact that coupling $\psi_{\pm}$ to gauge fields independently in two dimensions is not well-defined. To see this, we represent the gauge field as $$A_{\mu}=(g^{-1}\partial_{+}g,h^{-1}\partial_{-}h),$$ where $g,h\in U(N)$ and $x_{\pm}=x_1\pm x_2$, or equivalently denote it concisely as $(g,h)$. After a gauge transformation given by $t(x)\in U(N)$, it is found that $$(g,h)\rightarrow(gt^{-1},ht^{-1}),$$ which means that $(g,h)$ is gauge equivalent to $(gh^{-1},1)$ and $(1,hg^{-1})$ that couple  respectively to $\psi_{+}$ and $\psi_{-}$. Thus  it is appropriate to employ a gauge invariant regularization scheme in two dimensions, which may not be equivalent to the above one. In addition, there is also an advantage working with this gauge invariant regularization, namely all the renormalizable terms are all contained in $\mathcal{J}'[M]$ that can be calculated exactly. To evaluate $Z[M]$, we construct a series of infinitesimal axial gauge transformations still parametrized by $\tau\in[0,1]$, and accordingly define 
\begin{equation*}
\begin{split}
\psi_{+}(\tau)=\tilde{M}(\tau/2)\psi_{+},&\quad \bar{\psi}_{+}(\tau)=\bar{\psi}_{+}\tilde{M}(-\tau/2),\\
\psi_{-}(\tau)=\tilde{M}(-\tau/2)\psi_{-},&\quad \bar{\psi}_{-}(\tau)=\bar{\psi}_{-}\tilde{M}(\tau/2),
\end{split}
\end{equation*}
and
\begin{small}
\begin{multline*}
\mathcal{L}(\tau)=-\bar{\psi}_{+}(\tau)[k_{+}-iA_{+}(\tau)]\psi_{+}(\tau)+g\bar{\psi}_{-}(\tau)\tilde{M}(1-\tau)\psi_{+}(\tau)\\-\bar{\psi}_{-}(\tau)[k_{-}-iA_{-}(\tau)]\psi_{-}(\tau)+g\bar{\psi}_{+}(\tau)\tilde{M}^{-1}(1-\tau)\psi_{-}(\tau).
\end{multline*}
\end{small}
Similar to the previous case, accumulating infinitesimal axial gauge transformations given by $M(d\tau/2)$, we obtain the Jacobian determinant $\mathcal{J}'[M]$ for the finite transformation, $M(1/2)$. In stead of calculating $\mathcal{J}'[M]$ directly, we note that $$\mathcal{J}'[M]= Z_0^{-1}\int \mathcal{D}\psi\mathcal{D}\bar{\psi}\exp(-\int \bar{\psi} i\slashed D \psi)\equiv e^{-S[A]},$$ where $D_{\mu}$ is the covariant derivative with $A_{+}=M^{-1}\partial_{+}M$ and $A_{-}=0$. An explicit expression of $S[A]$ has been derived in Ref.\cite{Polyakov} by solving the equations, 
\begin{equation*}
\begin{split}
D_{\mu}J_{\mu}&=0\\
\epsilon_{\mu\nu}D_{\mu} J_{\nu}&=\frac{1}{4\pi}\epsilon_{\mu\nu}F_{\mu\nu},
\end{split}
\end{equation*} 
coming from the gauge invariant regularization scheme. It is exactly the well-known Wess-Zumino-Witten model \cite{Witten-bosonization},
\begin{multline}
S=\frac{1}{8\pi}\int_{S^2} d^2x\mathrm{tr}(M^{-1}\partial_\mu M)^2\\
\quad +\frac{i}{12\pi}\int_{D^3}\mathrm{tr}(\tilde{M}^{-1}d\tilde{M})^3.
\label{WZW-model}
\end{multline}
Actually all the renormalizable terms in the effective action have been contained in Eq.(\ref{WZW-model}). It is interesting to note that  as a byproduct of our theory, a rigorous derivation of the Wess-Zumino-Witten model of Eq.(\ref{WZW-model}) , which is an exactly solvable conformal field theory~\cite{Polyakov}, has been given as an exact result for $\mathcal{H}_{2d}^D$ for a given replicate number $N$. Also, it should be noted that only the coefficient of WZ term (the second one) in Eq.(\ref{WZW-model}) has a topological origin, and therefore is determined by the topological charge of $\mathcal{H}_{2d}^{D}$ in accord with Eq.(2) ($\nu=1$) in the main text. 

As a nontrivial application of Eq.(\ref{WZ-AIII}), there exists the term of Eq.(\ref{WZ-AIII}) with $\nu_{AIII}=2$ in the NL$\sigma$M of the quadratic Fermi point in the class AIII, $$\mathcal{H}_{AIII}(k)=(k_x^2-k_y^2)\sigma_1+2k_xk_y\sigma_2,$$ which is difficult to be derived directly. The interpolation of the quadratic Fermi point to doublet of $\sigma_1k_x+\sigma_2k_2$ is shown in Appendix \ref{AIII-Inter}. It is noted that Fermi point of this form can appear as the gapless modes on the boundary of a topological crystalline insulator \cite{Crystal-TI}.

\section{Real classes} 
It is proposed that a topological insulator or superconductor (TI/TSC) with bulk topological number $N\in \mathbb{Z}$ has a WZ term on the disordered boundary, while one with nontrivial $\in\mathbb{Z}_2$ has a $\mathbb{Z}_2$-$\theta$ term, which has been used to deduce the classification of TIs/TSCs \cite{Schnyder-classification}. Due to the boundary-bulk correspondence that the bulk topological number is equal to the total topological charge of boundary gapless modes \cite{TI-FS}, we may propose that the topological term in Eq.(\ref{Top-term}) for a Fermi point with topological charge $\nu\in \mathbb{Z}$ be just a WZ term at level $\nu$, while that for a Fermi point with nontrivial $\mathbb{Z}_2$ topological charge be just a $\mathbb{Z}_2$-$\theta$ term. More clearly this proposition is supported by a one-to-one correspondence of the classification of topological Fermi points in real classes to the distribution of WZ and $\mathbb{Z}_2$-$\theta$ terms. However, more rigorous derivation of the conclusion is still awaited, although for a $\mathbb{Z}_2$ Dirac point in the class AII has been derived numerically\cite{Ryu-Z2}.

Now let us derive the one-to-one correspondence of the classification of topological Fermi points in real classes to the distribution of WZ and $\mathbb{Z}_2$-$\theta$ terms directly. For the class $q$, due to the constraints from the corresponding symmetries, the one-point Hamiltonian is in the manifold $R_q$ called the $q$th classifying space, noting that the subscript $n$ of $R_n$ is an integer modulo 8. Under the symmetry-preserving randomness, the target space of the NL$\sigma$M is $T_{q}\approx R_{4-q}$, and if the dimension is $d$, the homotopy group is given by $$\pi_{d}(T_{q})\cong \pi_{0}(R_{4-q+d})$$ with $$\pi_{0}(R_{q})\cong \mathbb{Z}, \mathbb{Z}_2, \mathbb{Z}_2, 0, \mathbb{Z}, 0, 0, 0$$ for $q=0,\cdots,7$. There are two kinds of topological terms with discrete coefficients, consisting of WZ terms and $\mathbb{Z}_2$-$\theta$ terms. As an important result in Ref.\cite{Schnyder-classification}, the WZ terms are possible when $\pi_{d}(T_q)\cong 0$ and $\pi_{d+1}(T_{q})\cong \mathbb{Z}$, which are satisfied if $d+1-q\equiv 0 \mod 4$, while $\mathbb{Z}_2$-$\theta$ terms exist when $\pi_{d}(T_{q})\cong \mathbb{Z}_2$, correspondingly $4-q+d\equiv 1$ or $2 \mod 8$, which is encapsulated as
\begin{equation*}
d-q\equiv \begin{cases}
7 \mod 8, & \mathrm{WZ}\\
5\,\,\mathrm{or}\,\,6 \mod 8, & \mathbb{Z}_2\theta
\end{cases}.
\end{equation*}
On the other hand, the classification of a Fermi point in the $q$-th class in a $d$-dimensional $k$ space is given by  \cite{FS-classification,TI-FS}$$\pi_{0}(R_{q-(d+1)}).$$ Thus the $\mathbb{Z}$ topological charges appear when $q-(d+1)\equiv 0 \mod 4$, which is just the condition for WZ terms. And $\mathbb{Z}_2$ topological charges exist when $q-(d+1)\equiv 1 \mod 8$ or $q-(d+1)\equiv 2 \mod 8$, exactly the conditions for $\mathbb{Z}_2$-$\theta$ terms, which is easy to check. To conclude, this one-to-one correspondence of topological charges to WZ and $\mathbb{Z}_2$-$\theta$ terms strongly suggests that the topological term for a Fermi point with topological charge $\nu\in \mathbb{Z}$ be just a WZ term at level $\nu$, while that for a Fermi point with nontrivial $\mathbb{Z}_2$ topological charge be just a $\mathbb{Z}_2$-$\theta$ term, even though it is extremely hard to analytically derive these topological terms due to the complicated structures of the classifying spaces $R_{q}$.

\begin{acknowledgments}
We thank A. Altland for drawing our attention to the bosonization method in the 2$d$ AIII case. This work was supported by the GRF (Grant Nos. HKU173055/15P and HKU173051/14P) and the CRF (HKU8/11G) of Hong Kong.
\end{acknowledgments}

\appendix
\section{The interpolation for AIII models}\label{AIII-Inter}
Consider a quadratic Fermi point in the class AIII,
\begin{equation}
\mathcal{H}_{AIII}(k)=(k_x^2-k_y^2)\sigma_1+2k_xk_y\sigma_2. \label{2-AIII}
\end{equation}
It is straightforward to check that the Fermi point has topological charge $\nu_c=2$ by following the formula
\begin{equation}
\nu_c=\frac{1}{4\pi i}\mathrm{tr}\oint_C dl~\mathcal{H}^{-1}\partial_l\mathcal{H}(k), \label{chiral-charge}
\end{equation}
where $C$ is a loop enclosing the gapless point in the $k$ space. We now manage to interpolate continuously between 
\[
\mathcal{H}_I(k)=\mathcal{H}(k)\otimes (\tau_0+\tau_3)/2+\sigma_x\otimes(\tau_0-\tau_3)/2
\]
and
\[
\mathcal{H}_{II}(k)=(k_x\sigma_1+k_y\sigma_2)\otimes\tau_0,
\]
without breaking the chiral symmetry $\sigma_3$ and violating the asymptotic behavior, where $\mathcal{H}_{I}$ is just the Fermi point of Eq.(\ref{2-AIII}) with two trivial bands being added, and $\mathcal{H}_{II}$ is just a doublet of 2D Dirac points, with each consisting of a unit topological charge. We choose the loop $C$ in Eq.(\ref{chiral-charge}) as the unit circle parametrized by $\phi\in[0,2\pi)$, where $k_x=\cos\phi$ and $k_y=\sin\phi$. Then the homotopy is given by
\begin{multline}
\tilde{\mathcal{H}}(t,\phi)=\cos t\,\mathcal{H}_{I}(\phi)+\sin t\,\mathcal{H}_{II}(\phi)\\   +\lambda\sin t\cos t(\sigma_x+\sigma_y)\otimes\tau_x
\end{multline}
\begin{figure}
\centering
\includegraphics[scale=0.4]{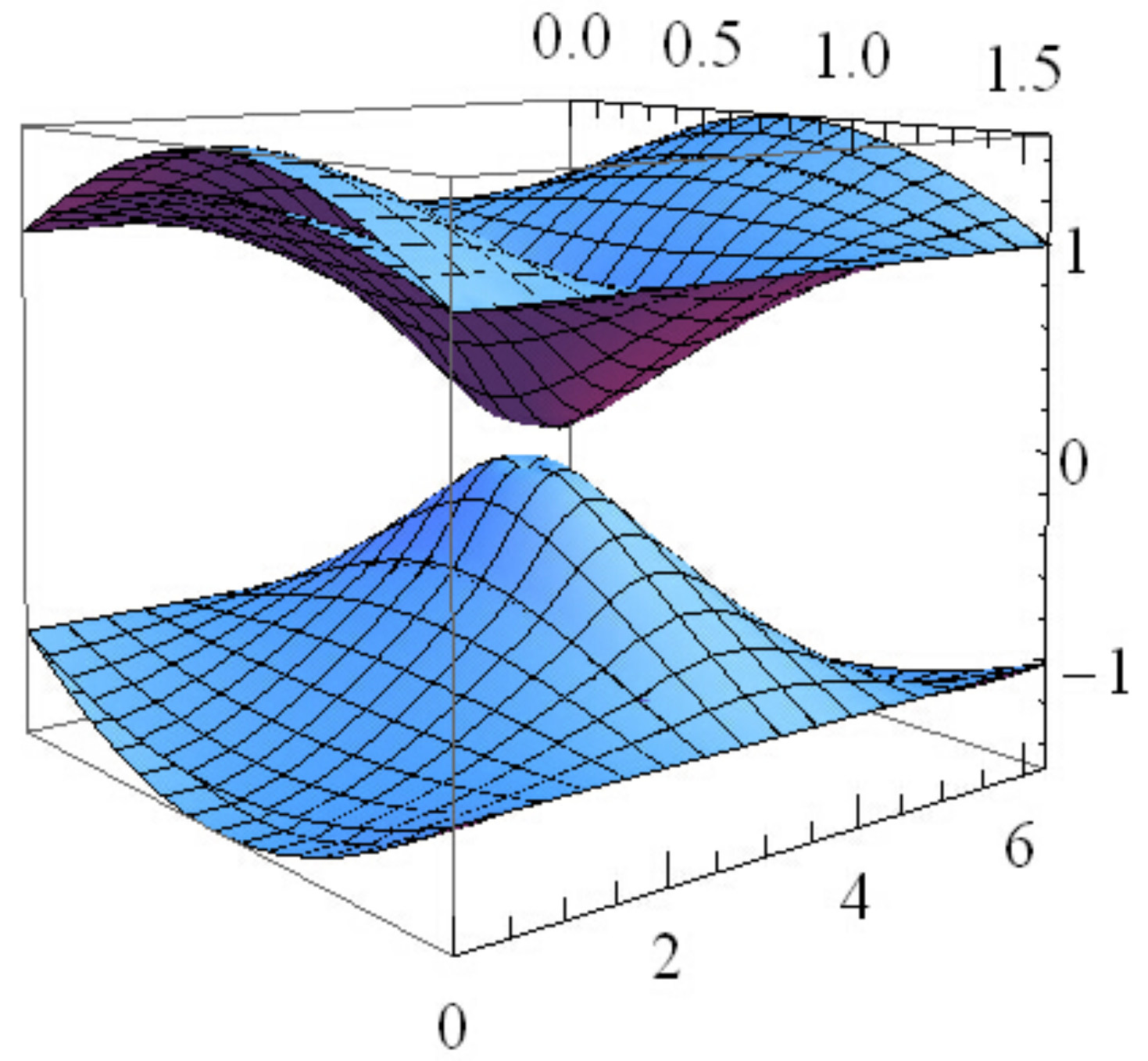}
\caption{The interpolation in the class AIII with $\lambda=0.2$. \label{Interpolation-AIII}}
\end{figure}
with $t\in[0,\pi/2]$. If $\lambda=0$, there is a gapless point at the center $(\pi/4,\pi)$ in the $t-\phi$ space, and through adding the perturbation term with small $\lambda$, the gap is opened, making the above equation a well-defined interpolation.Since no $\sigma_0$ and $\sigma_3$ are involved, the chiral symmetry is always preserved. The spectrum for $\lambda=0.2$ is shown in Fig.(\ref{Interpolation-AIII}). 

\section{The Interpolation for A-class Models}\label{A-Inter}
From Eq.(\ref{2-AIII}), it is reasonable to guess a quadratic model of topological charge 2 in the class A as
\begin{equation}
\mathcal{H}_{A}(k)=(k_x^2-k_y^2)\sigma_1+2k_xk_y\sigma_2+k_z\sigma_3.
\end{equation}
Choosing a unit sphere parametrized by $\theta\in[0,\pi]$ and $\phi\in[0,\pi)$, the Hamiltonian is restricted on the sphere as
\begin{equation}
\mathcal{H}_A(\theta,\phi)=\sin^2\theta\cos2\phi\sigma_x+\sin^2\theta\sin2\phi\sigma_y+\cos\theta\sigma_z.
\end{equation}
To see the topological charge is indeed 2, we replace $\sin^2\theta$ by $x\sin\theta+(1-x)\sin^2\theta$ in $\mathcal{H}_A(\theta,\phi)$. Through varying $x$ from $0$ to $1$, we see that $\mathcal{H}_A(\theta,\phi)$ is smoothly deformed to be
\begin{equation}
\mathcal{H}'_A(\theta,\phi)=\sin\theta\cos2\phi\sigma_1+\sin\theta\sin2\phi\sigma_2+\cos\theta\sigma_3
\end{equation}
without closing the gap on the sphere. It is obvious that $H'_A(\theta,\phi)$ has the winding number $2$ as the topological charge of the Fermi point. We now interpolate between
\begin{equation}
\mathcal{H}_I=\mathcal{H}'_A\otimes(\tau_0+\tau_3)/2+\sigma_3\otimes(\tau_0-\tau_3)/2
\end{equation}
and
\begin{equation}
\mathcal{H}_{II}=(\sin\theta\cos\phi\sigma_1+\sin\theta\sin\phi\sigma_2+\cos\theta\sigma_z)\otimes\tau_0/2.
\end{equation}
Based on the previous experience, we first make the leading-order interpolation
\begin{equation}
\tilde{\mathcal{H}}_0(t,\theta,\phi)=\cos t\mathcal{H}_I(\theta,\phi)+\sin t\mathcal{H}_{II}(\theta,\phi)
\end{equation}
with $t\in[0,\pi/2]$, which has eigen energies
\begin{equation}
\begin{split}
E^2=&9/2-\cos(2t)/2+4\cos\theta\sin t,\\ &5/2+3\cos(2t)/2\\
&+\sin(2t)(1+\cos\phi+\cos(2\theta)(1-\cos\phi)).
\end{split}
\end{equation}
It is found that there is only one gapless point at $(\theta_0=\pi/2,\phi_0=\pi,t_0=\pi/4+\arcsin(3/5)/2)$ in the whole space $S^2\times [0,\pi/2]$. We expand $\tilde{\mathcal{H}}_0$ around the gapless point and keep only up to linear terms of $\delta\theta$ and $\delta\phi$, and obtain
\begin{equation}
\begin{split}
\tilde{\mathcal{H}}_0=&(\cos(t_0/2) - \sin(t_0/2)) \sigma_1\otimes\tau_0 \\
&+ \delta\phi (2 \cos(t_0/2) - \sin(t_0/2)) \sigma_2\otimes\tau_0 \\
&+ (1 - (\cos(t_0/2) + \sin(t_0/2)) \delta\theta) \sigma_3\otimes\tau_0 \\
&+ \cos(t_0/2) \sigma_1\otimes\tau_3 - (1 + 
   \cos(t_0/2) \delta\theta) \sigma_3\otimes\tau_3\\
&+ 2 \cos(t_0/2) \delta\phi \sigma_2\otimes\tau_3 +\mathcal{O}(\delta\theta^2,\delta\phi^2,\delta\theta\delta\phi).
\end{split}
\end{equation}
And it is straightforward to check that the perturbation term
\begin{equation}
\Delta\tilde{\mathcal{H}}=2\lambda \sin t\cos t\sigma_2\otimes\sigma_3
\end{equation}
can open a gap
\begin{equation}
\Delta E=8\lambda/5
\end{equation}
with $0<\lambda\ll 1$. Thus we find that $\tilde{\mathcal{H}}(t,\theta,\phi)=\tilde{\mathcal{H}}_0+\Delta\tilde{\mathcal{H}}$ can achieve the interpolation, and through two steps of smooth deformations, the quadratic Fermi point with topological charge 2 can be transformed continuously to a doublet of Weyl Fermi points with the same chirality.

\end{document}